# Large Area Near-Field Thermophotovoltaics for Low Temperature Applications

Jennifer Selvidge[1], Ryan M. France[1], John Goldsmith[1], Parth Solanki[2], Myles A. Steiner[1], and Eric J. Tervo[1,2,3]*

[1]National Renewable Energy Laboratory, Golden, CO, 80401, USA
[2]Department of Electrical & Computer Engineering, University of Wisconsin-Madison, Madison, WI, 53706, USA
[3]Department of Mechanical Engineering, University of Wisconsin-Madison, Madison, WI 53706 USA

**Abstract:**
Thermophotovoltaics, devices that convert thermal infrared photons to electricity, offer a key pathway for a variety of critical renewable energy technologies including thermal energy storage, waste heat recovery, and direct solar-thermal power generation. However, conventional far-field devices struggle to generate reasonable powers at lower temperatures. Near-field thermophotovoltaics provide a pathway to substantially higher powers by leveraging photon tunneling effects. Here we present a large area near-field thermophotovoltaic device, created with an epitaxial co-fabrication approach, that consists of a self-supported 0.28 cm$^2$ emitter-cell pair with a 150 nm gap. The device generates 1.22 mW at 460 °C, a twenty-five-fold increase over the same cell measured in a far-field configuration. Furthermore, the near-field device demonstrates short circuit current densities greater than the far-field photocurrent limit at all the temperatures tested, confirming the role of photon tunneling effects in the performance enhancement. Modeling suggests several practical directions for cell improvements and further increases in power density. These results highlight the promise of near-field thermophotovoltaics, especially for low temperature applications.

*Author to whom correspondence should be addressed: tervo@wisc.edu

## 1. Introduction

Thermophotovoltaics (TPVs), solid-state heat engines that convert infrared photons to electricity, promise to play a key role in global decarbonization efforts.[1-4] They offer the benefits of thermoelectric generators, including low maintenance and modularity, but with the potential for improved performance. Recently, significant progress in the field has resulted in high TPV power conversion efficiencies,[5-10] outpacing thermoelectric efficiencies.[11] These high efficiencies makes TPVs relevant to a wide range of applications. However, although high power densities have been demonstrated with high temperature emitters,[7,12] low power densities result from low temperature emitters (see Figure S4), posing a key challenge for temperatures below about 750 °C with applications including industrial waste heat recapture[2,13] and solar thermal energy conversion.[14-18] Efforts to increase power density with low temperature emitters face a significant challenge: low temperature heat sources have lower spectral radiances that peak at lower photon energies.

Near-field TPV systems offer a pathway to higher power than is classically allowed (i.e., predicted by Planck's blackbody theory), making them uniquely suited for low temperature applications. Traditional TPV systems operate in the far-field, meaning that the distance between the TPV cell and the emitter ($d$) is substantially larger than the emitted photon wavelengths. However, in the near-field, the gap between emitter and cell is similar to or smaller than the photon wavelength, which can drastically increase radiative heat transfer due to photon tunneling.[19-24] Specifically, in the near-field, additional radiative transfer mechanisms mediated by frustrated modes (those normally trapped by total internal reflection)[25] and surface modes (surface plasmon-polaritons and surface phonon-polaritons)[26,27] can account for the vast majority of the radiative transfer. Illustrations of far-field and near-field radiative transfer mechanisms are presented in Figure 1(a) and 1(b), respectively. Figure 1(c) presents the resulting difference in energy transfer between the two cases, demonstrating a 12-fold increase in above-bandgap ($E_{photon} \geq 0.35$ eV) radiative heat transfer assuming a 50 nm gap, a 500 °C GaAs emitter, and a 20 °C InAs cell. The order of magnitude increase in available power is key to ensuring low temperature TPV viability.

Numerous theoretical studies emphasize the promise of integrated near-field TPV systems,[28-34] but experimental device demonstrations have been more limited by the difficulty in creating nano-scale gaps between emitter and cell.[19,20,35-40] Experimental studies frequently rely on approaches utilizing piezoelectric micro-positioners to demonstrate near-field TPV,[20,35-37] but such systems are not well suited for larger device sizes, as small changes in tilt across the sample can dramatically change the end to end gap size; the largest demonstrated emitter using a micro-positioner to date has a radius of 75 μm.[36] Thus, although these approaches often demonstrate high power densities, they do not address the need for higher total power. Others have employed physical spacers between emitter and cell[19,38,39] or used nano electromechanical system-based approaches[40]. Spacer-based approaches face additional difficulties due to parasitic conduction arising from physical contact between the emitter and cell, representing a key challenge for scalable approaches.[39,41,42] Creating large devices capable of generating high power densities at low temperatures remains a significant challenge.

Here we present a large-area near-field TPV approach designed for operation with medium grade heat sources. The device employs an epitaxially grown, self-supported GaAs-based emitter that sits directly on the cell to ensure strong evanescent coupling (Fig. 2(a)). The emitter is displaced from the cell using spacers consisting of long support posts fabricated by chemically etching deep into the emitter. The long geometry of the posts reduces parasitic thermal conductive losses where the cell and emitter are in physical contact.[38,41] The gap itself is subsequently defined by the selective etching of a thin heteroepitaxial layer everywhere but the post ends. Heteroepitaxy and selective etching ensure precise control of the gap size over a large area. Epitaxial growth and selective etching processes are extremely common in a wide variety of III-V semiconductor applications, assisting with scalability. The cell front metallization utilizes very thin grids, eliminating the need for complex alignment for the full device. These devices also require low bandgap cells that can operate effectively at near-ambient temperatures without the need for liquid nitrogen cooling. To that end, we use InAs cells with very low reverse bias saturation current compared to the state of the art. Using a co-designed III-V based emitter-cell pair yields a 1.22 mW output power at 460 °C and a nominal gap size of 150 nm from a 0.28 cm$^2$ active area. These results highlight the potential of near-field TPV for industrial applications and identify key areas for continued development.

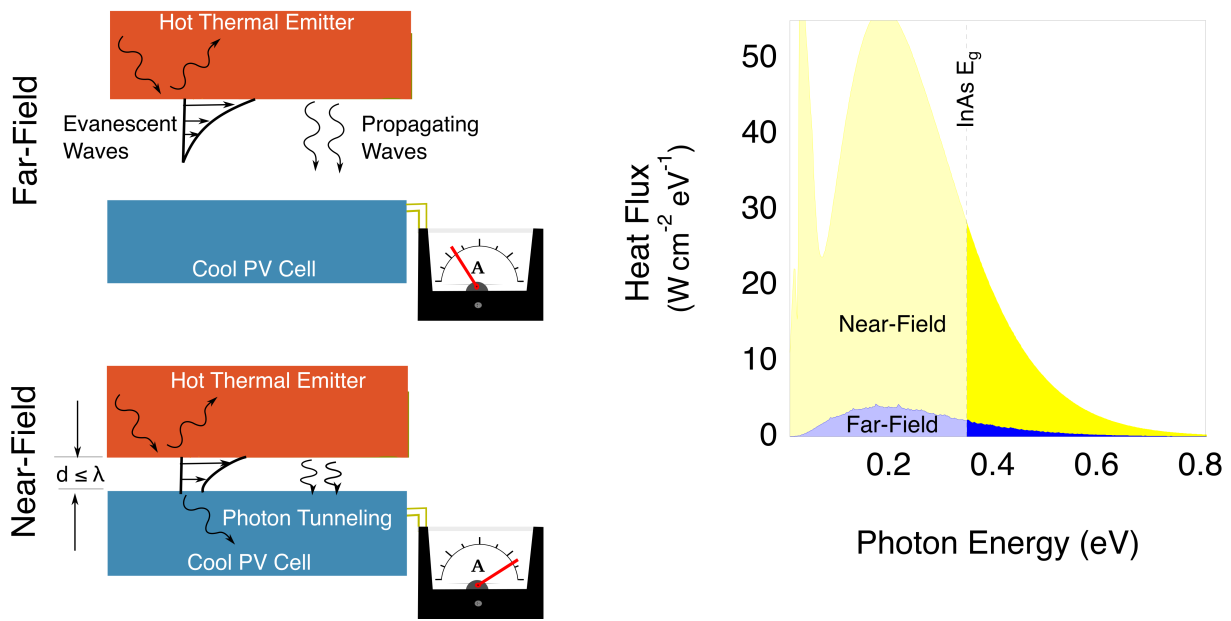

Figure 1. Schematic of radiative heat transfer mechanisms in (a) the far-field and (b) the near-field. In both cases, propagating waves transfer energy between the hot emitter and cool cell. In the near-field, photons normally trapped by total internal reflection and surface waves evanescently couple to the cool cell, greatly increasing the total radiative energy transfer. (c) Photon flux as a function of photon energy between a 500 °C GaAs emitter and a 20°C doped InAs cell spaced 50 nm apart (near-field) and hundreds of microns apart (far-field). Only above-bandgap energy can be converted into electrical energy, (dotted line denotes InAs bandgap at 300 K) so low bandgap materials with optimized cell designs play a key role in system design. The low energy peak arises from surface mode coupling.[41]

## 2. Results and Discussion

### 2.1 Fabrication

Our fabrication approach relies on industry-standard growth and processing techniques to improve both device and process scalability. As outlined in Figure 2(a), both the TPV cell and emitter are grown using metalorganic vapor phase epitaxy (MOVPE). The GaAs emitter and InAs cell are designed to be integrated together, and so are co-fabricated to ensure easy assembly (see Experimental Section and Fig. S1 and Fig. S2 for additional detail.) InAs[43] and GaAs[44] have similar refractive indices in the mid-IR, facilitating evanescent coupling of frustrated modes between the hot emitter and cool cell.

The emitter consists of a lattice-matched nominal 150 nm GaInP spacer grown on an n+ GaAs substrate, with the thickness of the spacer determined by the calibrated growth rate of the MOVPE reactor. After growth, the GaInP spacer is selectively wet etched everywhere except for the locations of the four supporting posts, which makes the GaInP spacer thickness define the size of the nanoscale gap. The underlying GaAs is deeply chemically etched in the area surrounding each of the four posts, ensuring a relatively long conduction path from the emitter to the cell through these posts. Full processing details are available in the experimental methods. The emitter posts have a cross-sectional area of approximately 250 $\mu m^2$ and a height of approximately 10-15 $\mu m$ (Fig. 2(b)) to reduce parasitic conduction to the cell. Assuming a cell temperature of 20 °C and an emitter temperature of 500 °C, approximately 19-26% of the total energy transferred to the cell occurs through conduction, as described in the Supplemental Information.

Following the fabrication of both emitter and cell, the emitter is placed directly on the cell in a clean environment to avoid particulates, as shown in Fig. 2(a). To illustrate the geometry, Figure 2(c) shows an SEM image of a larger-gap test structure consisting of a nominally 500 nm GaInP spacer, showing an actual gap size of 600 $\pm$ 50 nm. The SEM imaging details are available in the experimental methods. The difference between the measured and nominal gap sizes is most likely due to an error in the growth calibration, while the 50 nm uncertainty comes from the approximately 20-30 nm uncertainty in the accuracy of both the upper and lower edge locations at this magnification. Figure S5 presents a lower magnification SEM image of the same sample, showing that the ~600 nm gap is maintained over at least 100 microns laterally and likely over the entire 0.5 cm x 0.5 cm lateral dimensions. The evidence of charging from the incident beam on the emitter side but not on the cell side in Fig. S5 indicates that the two sides are not in electrical contact near the imaging area. The gap cannot be clearly resolved at lower magnifications.

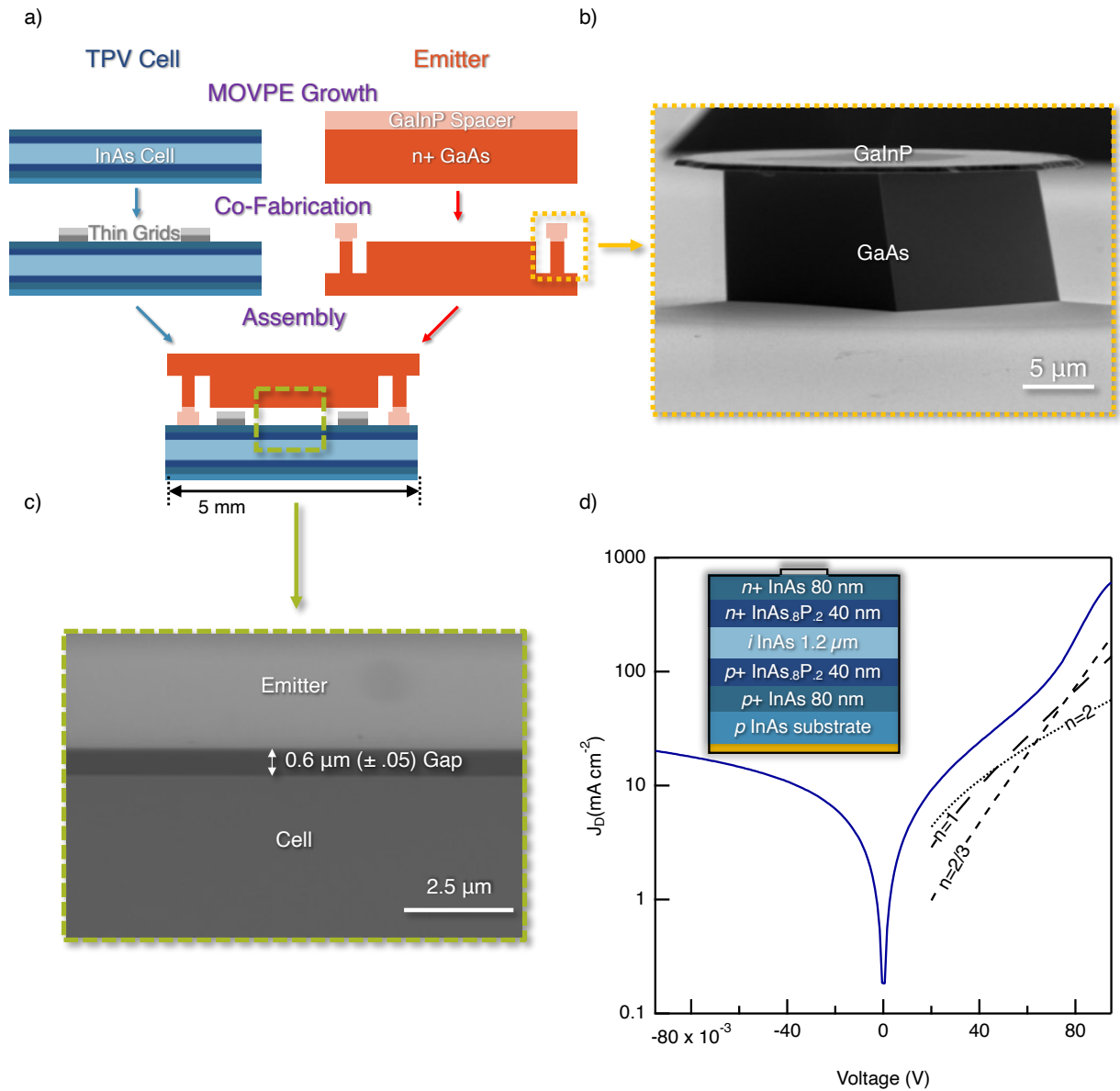

Figure 2. a) Major process flow steps for co-designed cell and emitter pair: (1) MOVPE growth of both emitter and cell, (2) complementary fabrication for emitter and cell, using thin grid geometry on the cell side and selective wet etching on the emitter side to create long thin posts, and (3) assembly of the emitter onto the cell. b) SEM image of an emitter post fabricated using the process in (a). The selective wet etching between the GaInP and GaAs leads to round spacer and smaller square post base. c) SEM image of an emitter and cell pair fabricated in using the process in (a) showing a 600 ± 50 nm gap size. d) Dark current-voltage curve for the TPV cell, showing a reverse bias current of approximately 20 mA cm$^{-2}$. The dotted and dashed lines are guides to the eye for n=1, 2 and 2/3 ideality behaviors; the curvature in the lines near V = 0 V occurs as the exponential term in the diode equation approaches 1. The inset presents a schematic of the full cell structure.

InAs has a bandgap of approximately 0.35 eV near room temperature and so is an appropriate cell material for low temperature sources.[45–47] Assuming a 500 °C emitter and a far-field device configuration, the total above-bandgap radiation intensity for an InAs cell operating at room temperature is 255 mW cm$^{-2}$. By contrast, the above-bandgap radiation intensity is only 5 mW cm$^{-2}$ for an In$_{.47}$Ga$_{.53}$As cell ($E_g = 0.74$ eV) with the same emitter. Figure S3 shows the blackbody spectrum at various temperatures and indicates the bandgap energies for InAs and In$_{.47}$Ga$_{.53}$As. An n-i-p double heterostructure (Fig 2(a)) minimizes Auger-Meitner recombination, a key limiting factor in low bandgap devices and high current devices.[48] The cell front metal electrode is designed with unusually short but wide grid fingers (0.1 $\mu$m x 100 $\mu$m) with a grid pitch of 400 $\mu$m, in order to lay directly beneath the emitter without the need for complex alignment or bonding equipment. Fig. 2(d) shows the current density as a function of voltage for the InAs cell at a nominal temperature of 10 °C and measured in the dark. The reverse current is approximately 20 mA cm$^{-2}$ at -0.09 V which compares favorably with previous reports of reverse bias saturation current densities on the order of a few tens to a few hundreds of mA cm$^{-2}$ for InAs TPV cells operating near room temperature.[49–53] Minimizing reverse bias current is particularly important as open circuit voltage, a key figure of merit for photovoltaic devices, decreases rapidly with increasing reverse bias saturation current.[54–56] The reverse bias current in these devices includes contributions from Shockley-Read-Hall (SRH) recombination in the depletion region (corresponding to n=2), bulk and interfacial SRH recombination in the quasi-neutral regions (corresponding to n=1), and Auger-Meitner recombination in the depletion region (corresponding to n=2/3).

**2.2 Testing**

To evaluate the near-field device, we designed a custom measurement facility. As shown in Figure 3(a), the heat source—a heated block of stainless-steel bar—is in physical contact with the emitter to ensure good conductive heat transfer between the two. The emitter temperature was measured at the indicated thermocouple (TC) position. The cell sits on a copper post mounted to a water-cooled plate to dissipate heat from the cell. Additional details are shown in Figure S3.

Electrical characterization is performed on one InAs cell with two emitters with different gap sizes. One emitter is intended for near-field operation and has a nominal 150 nm gap, while a second emitter is fabricated in the same manner but with an approximate 10-15 $\mu$m gap to compare far-field performance, exceeding the photon wavelengths above the InAs bandgap by three to five-fold. As near-field effects occur when $d \leq \lambda$, where d is the gap size and $\lambda$ is the photon wavelength, this should well-approximate the far-field. Testing the same cell with two different emitter gaps eliminates cell-to-cell variation as a factor in the current-voltage characteristics. The effective area of the cell is 0.53 cm$^2$. Due to the proximity of the emitter to the cell in both the near-field and far-field cases, we use the full emitter areas of 0.28 cm$^2$ in the near field case and 0.35 cm$^2$ in the far-field case (including the front metallization) as the active areas. The size discrepancy between the three areas (emitters and cell) arises from imprecision in cleaving. The thin grids likely reflect some incident radiation back to the emitter, but may also facilitate additional evanescent coupling via surface waves,[57] making their behavior complex.

The experimental near-field TPV data demonstrate a significant increase in the short circuit current, and thus the generated power, as compared to the far-field data, shown in Fig. 4(b-d).[35–40] Comparing the cell at the two highest emitter temperatures, 460 °C and 455 °C in the near- and far-fields respectively, the near-field short circuit current is an order of magnitude greater than the far-field current. The measured near-field short circuit current density is also significantly larger than the modeled far-field limit that assumes an internal quantum efficiency (IQE) of 100%, as shown in Figure 3(d), which unambiguously confirms the contributions of near-field radiative heat transfer to the photocurrent.

The data show exponential growth in the both the short circuit current density, Fig. 3(d), and the peak power, Fig. 3(e), with increasing emitter temperature, consistent with prior near-field TPV work.[35,36,38–40] The peak power is 0.05 mW at 455 °C in the far-field case and 1.22 mW at 460 °C in the near-field case, a 25-fold increase that is consistent with previous near-field reporting for gap sizes on the order of 100-200 nm.[35,36,38]

The near-field and far-field current-voltage curves have different shapes; specifically, in the near-field case the curve is nearly linear in the power quadrant. The linear behavior can likely be predominantly attributed to a series resistance and increased cell temperature. Linear current-voltage behavior in the power quadrant is frequently reported for experimental low-bandgap InAs TPV operating near room temperature in both the near-field[35] and far-field[49] and represents an area of active research. Because the voltage of the InAs device is low, series resistance impacts a large portion of the current-voltage curve. To better demonstrate the importance of these effects on low band-gap cells, Figures S6 – S8 present the impacts of series resistance and increasing temperature on the current voltage behavior and dark current behavior of low bandgap cells. The unusual curvature near 0.07 V in the near-field case suggests a possible non-linear resistance of unclear origin.

The open circuit voltage increases with short circuit current density to a lesser degree for near-field compared to far-field operation (Fig. 3(f)). Furthermore, at fixed $J_{sc}$, the far-field $V_{oc}$ is much greater than the near-field $V_{oc}$. $V_{oc}$ droop with increasing cell temperature is well documented in both TPV[5,49] specifically and concentrator photovoltaics[54–56] more broadly as increased temperatures lead to increases in reverse bias saturation currents, increases in series resistances, and decreases in shunt resistances, all of which degrade cell performance. The actual junction temperature is likely higher than the temperature measured at the cell-side thermocouple in the testing apparatus (Fig. 3a). Indeed, while the measurements were all conducted with a nominal cell temperature of approximately -10 °C, the modeled cell temperature (assuming a 50% IQE) in the near field case is approximately 50 °C (see Fig. 4a). Further, our modeling suggests significant below-bandgap radiative heat transfer in the near-field case (see Fig. S9 for additional details) consistent with cell heating. These results emphasize the importance of optimized system design for near-field TPV, with special consideration given to cooling requirements.

The near-field TPV results represent a high total output power using emitter temperatures at or below 800 °C. While there is still room for improvement, discussed in the next section, the power exceeds that of previous reporting for InAs based cells for both far-field configurations (highest single cell power = 0.057 mW at 500 °C)[50] and near-field configurations (highest cell power = 0.007 mW at 460 °C)[37] operating at temperatures near 500 °C. In addition, the near-field configuration leads to 0.04 mW of output power at 320 °C, comparable in magnitude to the 0.05 mW measured at 455 °C in the far-field case, emphasizing near-field TPV's unique suitability for low temperature heat sources.

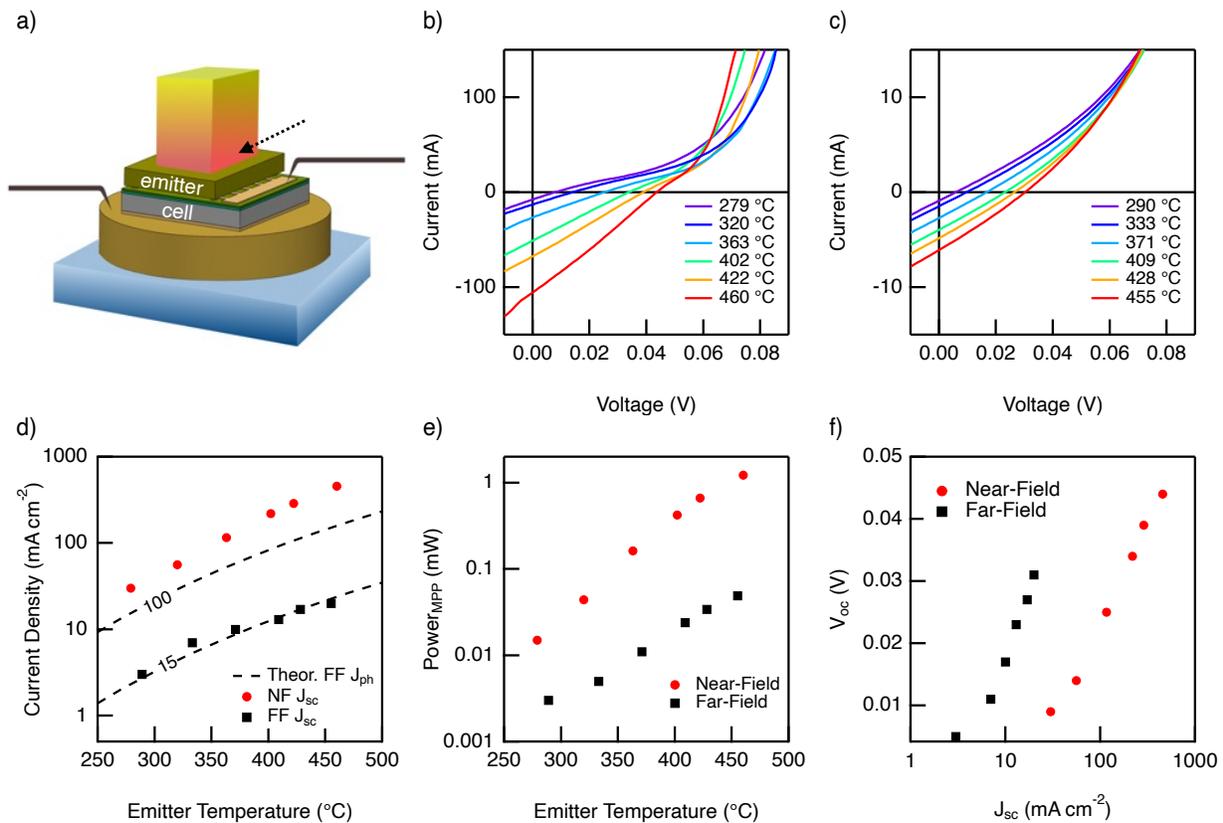

Figure 3. a) Schematic of near-field TPV test facility. Current-voltage behavior at various emitter temperatures for the cell in b) near-field and c) far-field configurations (note the 10X difference in scale on the current axes). d) Comparison of the measured short circuit current density as a function of emitter temperature in the near- and far-field configurations and theoretical photocurrents assuming an internal quantum efficiency of 100% and a 150 nm gap in the near-field case. The measured near-field data exceed the theoretical far-field limit. Note that while the measured $J_{sc}$ can be affected by the series and shunt resistances, the theoretical photocurrent is not. e) Comparison of generated power at the maximum power point as a function of emitter temperature in the near- and far-field cases, demonstrating more than an order of magnitude improvement in the near-field case. f) Open circuit voltage as a function of short circuit current density in the two cases, suggesting cell heating in the near-field case.

## 2.3 Pathways to Higher Power

There are several possible areas for system level and device level design improvements in ongoing work. As the measurements in Figure 3(f) suggested possible cell heating, we first consider the impact of cell temperature on our existing devices (Fig. 4(a)). We assume an emitter temperature of 460° C and an IQE of 50% leading to a modeled photocurrent of 606 mA cm$^{-2}$. We model the near-field TPV cell using a three-diode model, including Auger-Meitner recombination. For the three-diode model, the Auger-Meitner dark current varies with temperature according to Equation 1:

$$J_{0,aug} = C_1 T^{4.5} \exp\left(\frac{-E_g}{2/3\, k_b T}\right)$$

where $T$ is the cell temperature, $E_g$ is the temperature dependent bandgap, $k_b$ is the Boltzmann constant, and $C_1$ is a proportionality constant. The proportionality constants for $J_{0,1}$, $J_{0,2}$, and $J_{0,aug}$ can be extracted from the dark current-voltage measurement for the cell (Fig. 2(d)). Using these assumptions, we model an approximate cell temperature of 50 °C and an approximate series resistance of 65 mΩ•cm$^2$. While these values will change for different assumed IQE values, the trends in cell performance hold regardless. That is, as cell temperature increases, the open circuit voltage and the short circuit current densities decrease. As noted previously, the reduction in short circuit current density with increasing temperature (assuming non-zero series resistance) or increasing series resistance is prevalent in low bandgap cells (see Supplemental Figures 4, 5, and 6) operating near ambient temperatures. Series resistance reduces the maximum voltage at which the $J \cong J_{ph}$. For example, for a cell operating with a 1 A cm$^{-2}$ photocurrent, a series resistance on the order of 50 mΩ•cm$^2$ reduces the maximum voltage where $J \cong J_{ph}$ by approximately 0.1 V. For a large or moderate bandgap cell with $V_{oc} > 0.1$ V the maximum voltage at which $J \cong J_{ph}$ will be greater than 0 V and so $J_{sc} \cong J_{ph}$. In contrast, for a low bandgap cell, if $V_{oc} < 0.1$ V, the maximum voltage where $J \cong J_{ph}$ will be less than 0 V so $J_{sc} \neq J_{ph}$. The reduction in both open circuit voltage and short circuit current density with increasing temperature can lead to a significant reduction in output power. Our modeling suggests that lowering the cell temperature to room temperature should result in an approximate four-fold increase in power at an emitter temperature of 460 °C.

In addition to system level improvements, improvements to the device itself should also yield substantial increases in power. Figure 4(b) shows the anticipated output power improvements from a series of cumulative cell design improvements, assuming the cell is consistently cooled to 20 °C. Reducing the estimated series resistance (65 mΩ cm$^{-2}$) by 80% to 13 mΩ cm$^{-2}$ nearly doubles the output power at 500 °C. The series resistance likely arises due to the wide, flat front electrode geometry. Counter-sinking the front electrodes into the emitter would enable improved concentrator cell grid geometries. Removing the highly doped *n+* InAs contact layer should have two beneficial effects: 1) the photocurrent will increase by eliminating parasitic band-to-band and free carrier absorption, and 2) the IQE may improve, regardless of the photocurrent, consistent with recent modeling.[48] The heavily doped contact layer currently serves as a potent source of Auger-Meitner recombination as the depletion region width on the n-side of the device will be much wider than on the p-side due to differential doping (5x10$^{17}$ cm$^{-3}$ and 1x10$^{19}$ cm$^{-3}$, respectively) meaning minority carriers may be sufficiently close to the contact layer to enable Auger-Meitner recombination there. Here, we assume an improved IQE of 75% following contact layer removal in line with prior modeling.[48]

Power could be further increased by using a thin film geometry with a highly reflective Au back reflector, which could be obtained by removing the cell from the substrate via chemical etching or epitaxial liftoff. We modelled a thin film geometry for our current nominal gap size (150 nm) and one for a 50 nm gap size. A thin film geometry should result in a significant increase in photocurrent, because light will pass multiple times through the intrinsic region as it bounces between the reflector and the surface, as well as a significant reduction in sub-bandgap absorption (Fig S9(b)). In a thin-film geometry with a highly effective back reflector, most of the sub-bandgap light is reflected to the emitter itself rather than being absorbed and heating the cell. As such, the reduction in sub-bandgap absorption associated with a thin film geometry should also lessen cooling requirements at a given cell temperature.

Finally, near-field TPV offers a unique opportunity to improve cell output power simply by reducing the gap size. A 50 nm gap size further increases the photocurrent by enhancing the evanescent coupling effects between the emitter and the cell leading to an anticipated power density greater than 200 mW cm$^{-2}$ at 500 °C.

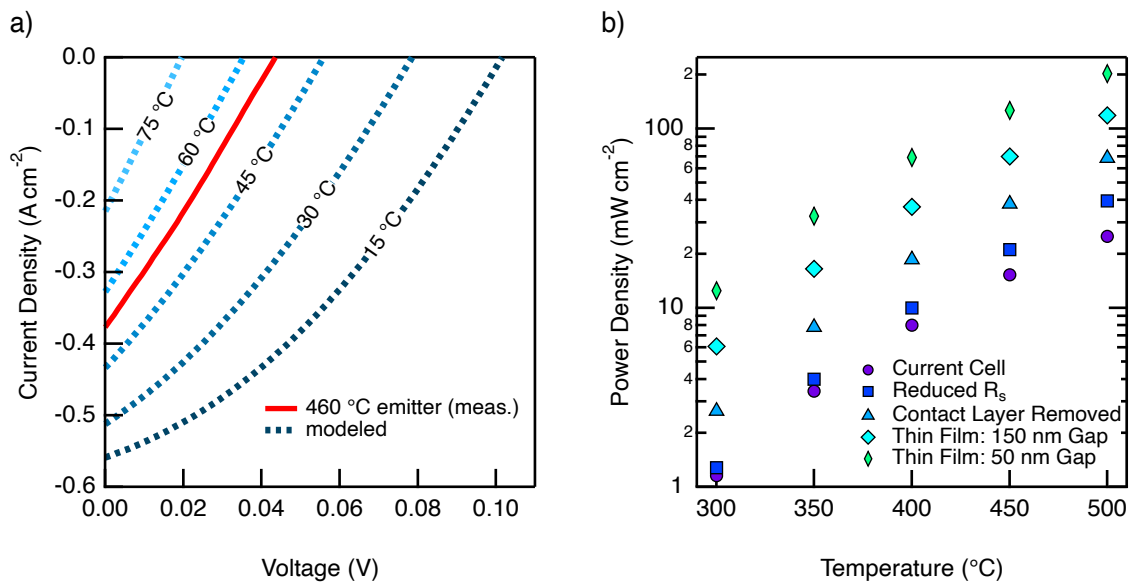

Figure 4. a) Modeled current-voltage curves assuming a three-diode model, including an Auger-Meitner specific dark current as a function of cell temperature and measured J-V with a 460 °C assuming a 0.606 A cm$^{-2}$ photocurrent. With a non-zero series resistance (assuming 65 mΩ cm$^{-2}$), both open circuit voltage and short circuit current density magnitude decrease with increasing temperature. b) Modeled power density at maximum power point as a function of temperature assuming various cell and device improvements.

## 3. Conclusions

In summary, we present a large area near-field thermophotovoltaic device capable of generating 1.2 mW of power (4.4 mW cm$^{-2}$) at 460 °C. This device demonstrates short circuit current densities exceeding the far-field photocurrent limit across the examined temperature range, confirming super-Planckian behavior. The total power output represents a 25-fold increase over the same device in a far-field configuration and a nearly three orders-of-magnitude increase over prior near-field work near 500 °C. The device demonstrates comparable power densities to prior positioner-based approaches,[36] but has the potential to be scaled towards larger sizes through increased device area or by tiling devices into a module. We rely on state-of-the-art InAs-based cells designed to operate near ambient temperatures with reverse dark current densities of 21 mA cm$^{-2}$ at 10 °C. Our co-designed emitters, grown using standard epitaxial growth techniques, facilitate device sizes more than an order of magnitude higher than previous experimental near-field reports at 0.28 cm$^2$. We also described attainable strategies on both the system and cell design side to improve device power to approximately 200 mW cm$^{-2}$ at 500 °C. These results strongly suggest that near-field TPV could pave the way for practical low temperature TPV applications.

## Acknowledgements

This work was authored in part by the National Renewable Energy Laboratory, operated by Alliance for Sustainable Energy, LLC, for the U.S. Department of Energy (DOE) under Contract No. DE-AC36-08GO28308. This work was supported in part by the Laboratory Directed Research and Development (LDRD) Program at NREL. Additional support for this research was provided by the University of Wisconsin – Madison Office of the Vice Chancellor for Research with funding from the Wisconsin Alumni Research Foundation. The views expressed in the article do not necessarily represent the views of the DOE or the U.S. Government. The U.S. Government retains and the publisher, by accepting the article for publication, acknowledges that the U.S. Government retains a nonexclusive, paid-up, irrevocable, worldwide license to publish or reproduce the published form of this work, or allow others to do so, for U.S. Government purposes.

# Supplementary Information

*Experimental Methods*

**Growth**

Both the InAs cell and GaAs emitter were grown using atmospheric pressure metal organic vapor phase epitaxy using standard precursors. The InAs cell was grown on a Zn-doped (001) InAs substrate (Fig. 2d). First, an 80 nm InAs:Zn nucleation layer was grown, followed by a strained 40 nm $InAs_{0.8}P_{0.2}$:Zn cladding layer, the undoped 1.2 um InAs absorbing layer, a $InAs_{0.8}P_{0.2}$:Si cladding layer, and a 80 nm InAs:Si contact layer. All layers were grown at a growth temperature of 550 °C and growth rate of 3 $\mu$m/hr. The composition calibration of InAsP was determined using X-ray diffraction, and no misfit dislocations were observed in Nomarski phase contrast microscopy. The doping densities in the cladding layers were nominally $1\times10^{18}$ $cm^{-3}$ Si and Zn, determined using secondary ion mass spectroscopy on calibration samples. The emitter was grown at 650°C on a Si-doped (001) GaAs substrate miscut 6° towards (111)A. A 200 nm GaAs nucleation layer was grown followed by a $Ga_{0.51}In_{0.49}P$ layer, the thickness of which determined the gap spacing.

**Processing**

Following growth, both the cell and emitter were cleaned using acetone and isopropanol alcohol to remove any dust particles. For the InAs back contact, we spun Shipley 1818 (S1818) photoresist on front side of the wafer and bake for 5 minutes at 100 °C. The backside of the InAs was etched in a 2:1:10 volumetric mixture of ammonium hydroxide, hydrogen peroxide, and deionized water for 1 minute. Following this, approximately 2 $\mu$m of Au was electroplated on the backside of the wafer to form the back contact. Figure S1 presents the co-fabrication front side process for our InAs cell and Figure S2 presents the co-fabrication for the GaAs emitter.

For the InAs cell:

1. S1818 was spun onto the front side of the cell and pre-baked for 5 minutes at 100 °C.
2. The S1818 was patterned for the frontside metallization using contact photolithography. Following the exposure, the sample was immersed in chlorobenzene for five minutes and blown dry with $N_2$. The sample was then developed using Microposit MF-351 developer (MF-351).
3. A 1 nm Pd/50 nm Ge/50 nm Ag front contact stack was evaporated using electron beam evaporation to facilitate contact to n+ InAs.
4. The S1818 and excess metal was lifted off via sonication in acetone.

5. The contacts were baked to form a PdGe intermetallic compound for 48 hours at 160 °C.
6. S1818 was spun on the front side of the cell, the S1818 was pre-baked at 100 °C for five minutes. The cell was then cleaved to the desired size.
7. The cleaved cell sidewalls were etched in a mixture of 1 mL of sulfuric acid, 8 mL of hydrogen peroxide, and 80 mL of water for 4.5 minutes.
8. The S1818 was removed via sonication in acetone.

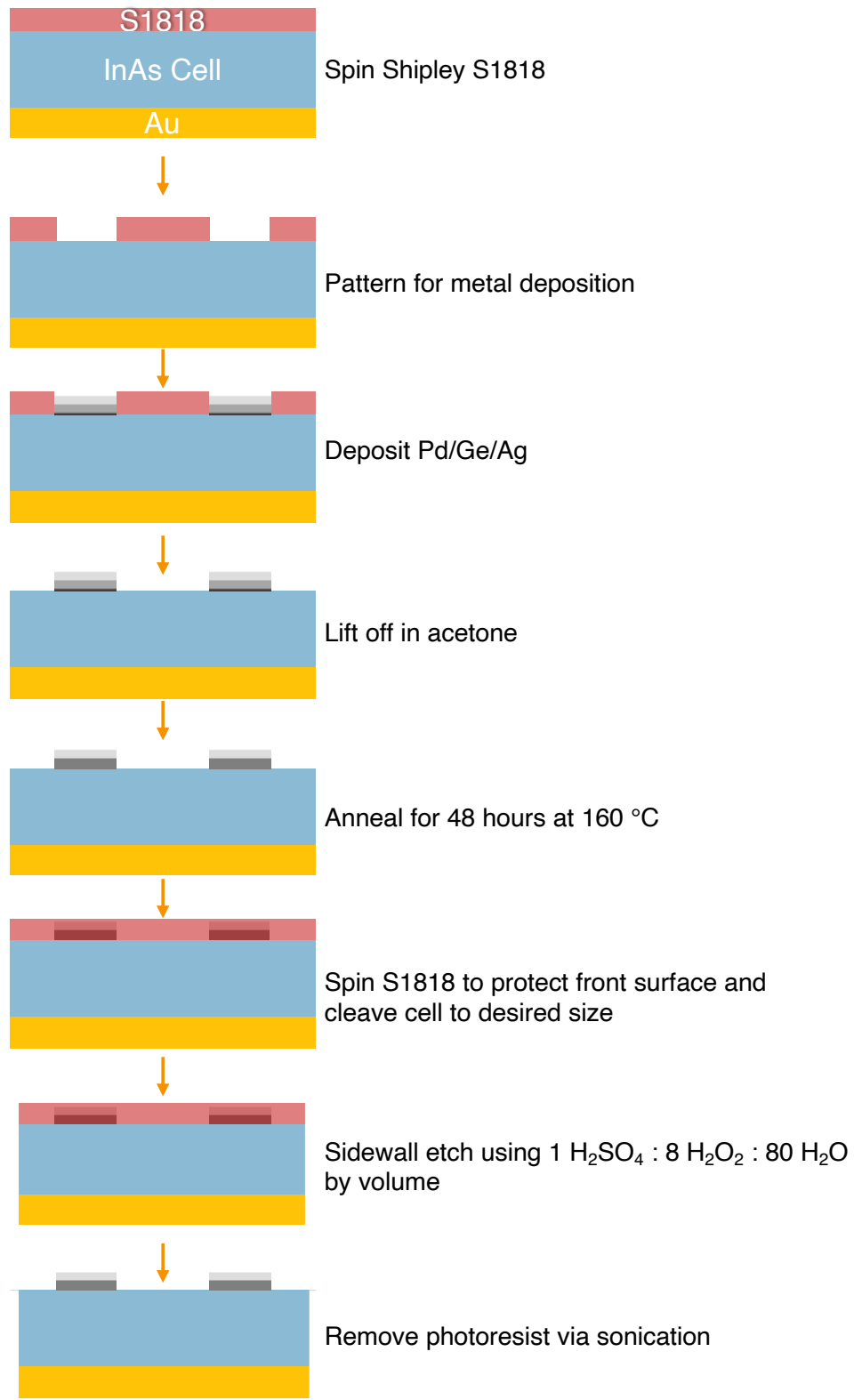

Supplemental Figure 1. Schematic of co-fabrication processing for the frontside of the InAs cell.

For the GaAs emitter:

1. S1818 was spun on the front side of the emitter and pre-baked at 100 °C.
2. The S1818 was patterned using contact photolithography and developed using MF-351 to create spacer posts surrounded by deep trenches.
3. The posts and trenches were created using selective wet etching of the GaInP and GaAs. The GaInP was etched using hydrochloric acid for 10 seconds and the underlying GaAs is etched using a 2:1:10 volumetric mixture of ammonium hydroxide: hydrogen peroxide: de-ionized water for 10 minutes to the desired depth (approx. 10 $\mu$m).
4. The S1818 was removed using acetone.
5. S1818 was spun on the front side of the emitter and pre-baked at 100 °C.
6. The S1818 was patterned using contact photolithography and developedusing MF-351 to protect the GaInP spacer layer on the posts and remove it elsewhere.
7. The GaInP was removed using a 5:1 volumetric mixture of phosphoric acid: hydrochloric acid. This mixture prevents undercutting of the spacer material on the posts.
8. The remaining S1818 was removed using acetone followed by a 5 min oxygen plasma reactive ion etching clean.

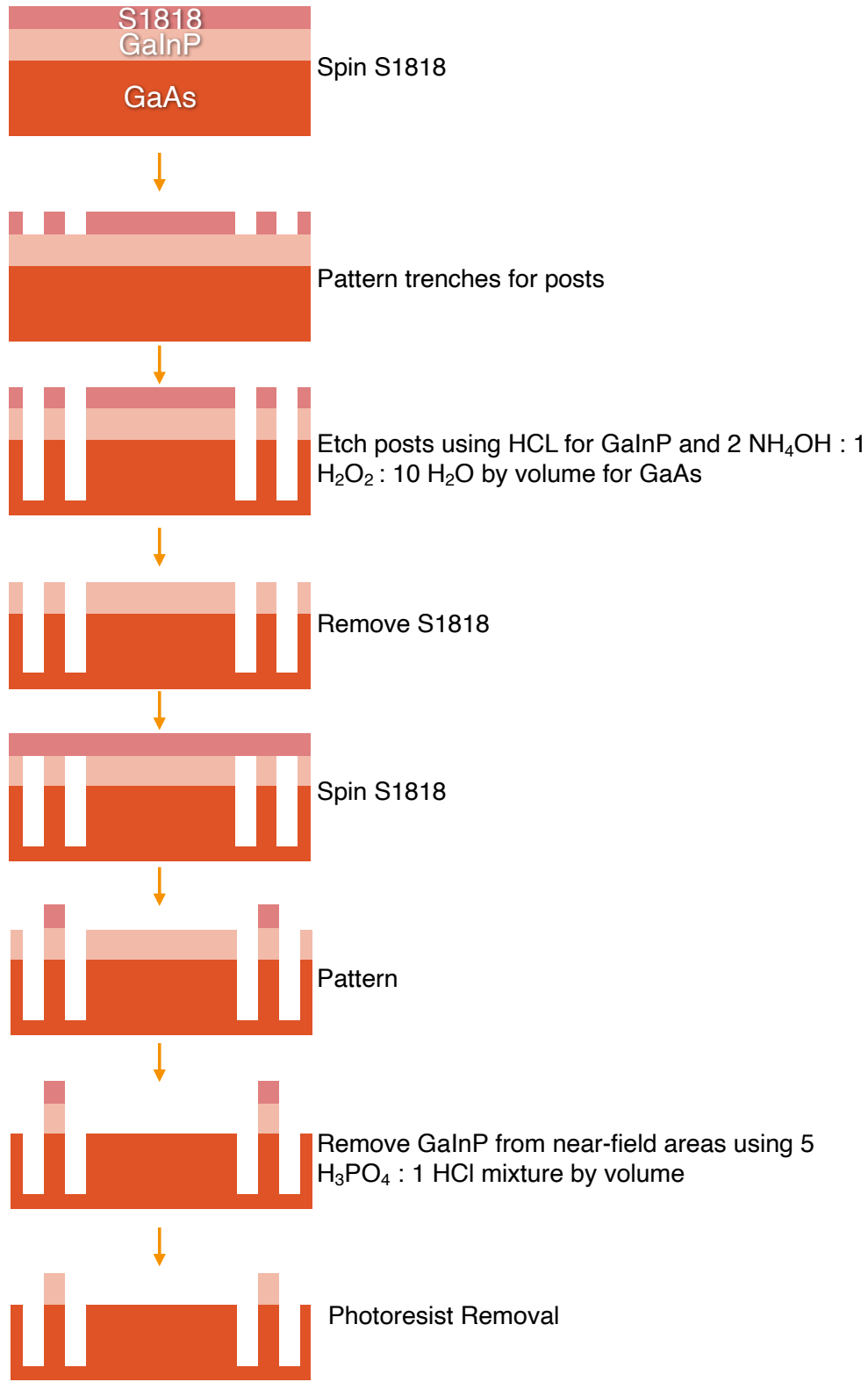

Supplemental Figure 2. Schematic of co-fabrication processing for the frontside of the InAs cell.

Following the completion of the fabrication of both components, the emitter was placed on the cell in the cleanroom and held together until being placed on the measurement platform for current-voltage measurements.

**Electrical Measurements**

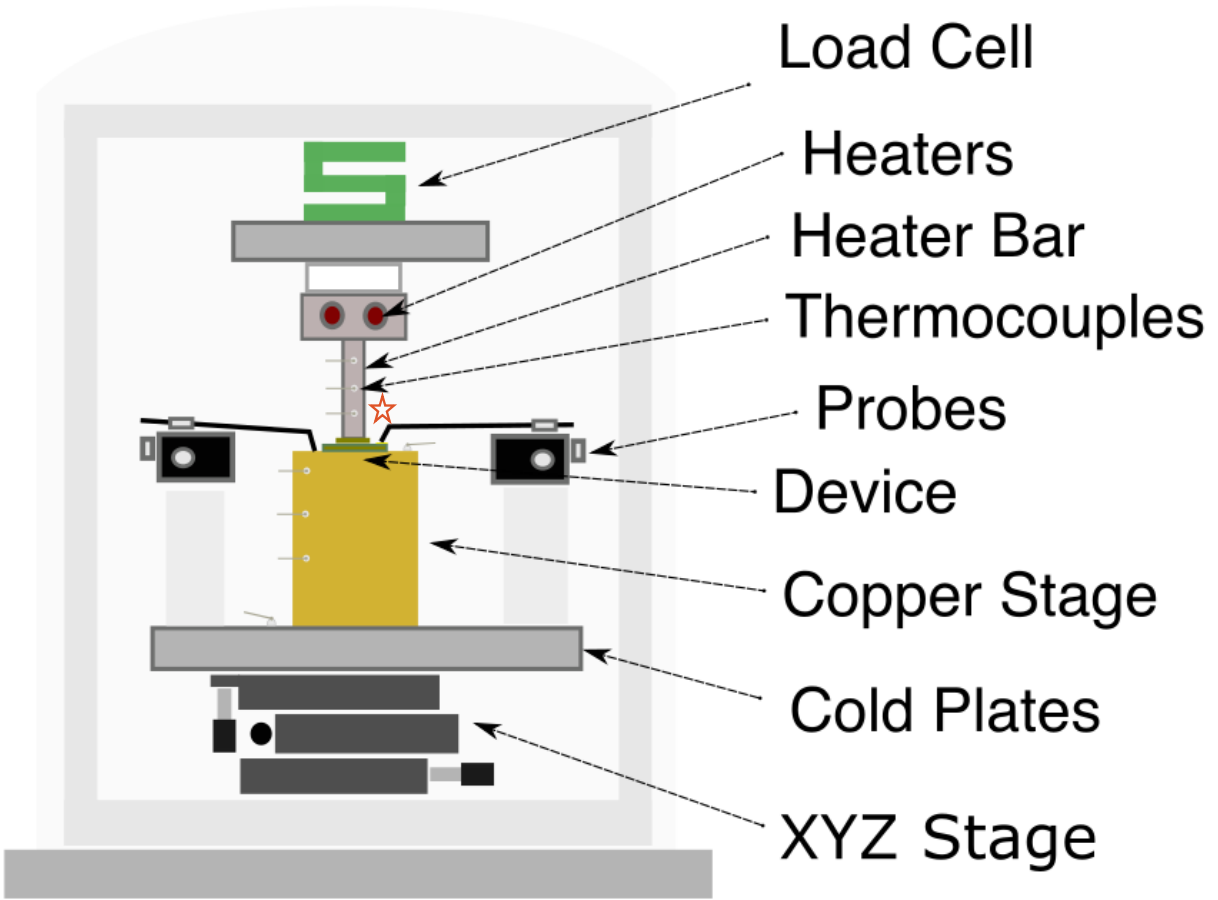

Supplemental Figure 3. Schematic of near-field TPV test facility.

Figure S3 presents a schematic of the measurement set-up used to perform illuminated current voltage measurements. On the cell side, the device sat on a piece of indium foil on a silver coated AlN puck on a copper stage atop a water-cooled plate. Thermal grease was applied at the interfaces between the AlN and the copper stage and the copper stage and the water cooled plate to ensure good thermal contact. The indium foil provides thermal and electrical contact between the cell and the silver coated AlN puck. The AlN puck ensures the system is eclectically isolated and, in conjunction with the copper stage, helps to facilitate thermal transport away from the cell. The water-cooled plate was cooled such that the temperature measured by the thermocouple mounted to the top of AlN puck reads approximately 263 K. Four-wire electrical measurements are made by making two contacts to the cell busbar (front) and two contacts to the indium foil (back). The measurements were conducted sweeping from high to low voltage from a voltage of 0.1 V to -0.02 V in .001 V steps. The sweep range in the near-field case was reduced to prevent possible cell damage at large negative currents.

On the emitter side, two cartridge heaters capable of reaching 810 K were mounted inside a stainless-steel heater bar and used as the heat source. Thermocouples measured the temperature at various points along the heater bar. The temperature measured at the thermocouple marked by the red star is taken as the emitter temperature. To ensure that the emitter was in good thermal contact with the heater bar, the XYZ stage was raised until the load cell measured a reduction in the tensile load on the topside of apparatus of 0.1 - 0.4 lbs. The load cell was separated from the heater bar by a second water cooled plate to ensure it was not damaged by the high temperature components. To conduct temperature dependent current-voltage measurements, the system was evacuated to a pressure of approximately 1e-3 mbar.

**SEM Imaging**

The SEM image in Figure 2(c) was acquired using an Everhart-Thornley detector (ETD) in secondary electron mode with 5.0 kV excitation voltage on a ThermoFisher Nova 630 SEM.

The SEM image in Figure 2(d) was acquired using an (ETD) in secondary electron mode with an accelerating voltage of 3.0 kV on a Hitachi S-4800 SEM. To create the sample for imaging, the emitter and cell were assembled as above and held together by painting S1818 photoresist on the edges other than the edge of interest. The sample and S1818 were cured in an oven to ensure that the S1818 was dry and mechanically stable. The sample was then mounted in cross section for imaging.

*Computational Methods*

**Thermal conduction calculations**

To estimate the amount of heat transferred from the emitter to the cell through the supporting posts, we solved Fourier's law

$$Q = -\kappa A \frac{dT}{dz} \quad (1)$$

where $Q$ is the heat transfer rate, $\kappa$ is the temperature-dependent thermal conductivity of the post, $A$ is the post's cross-sectional area, $T$ is the local temperature, and $z$ is the coordinate along the post length. This assumes one-dimensional heat transfer. The GaInP layer is neglected as this is thin (150 nm) compared to the entire post length (about 10 µm). The temperature-dependent thermal conductivity of the n-type GaAs with dopant density of about 2×10$^{18}$ cm$^{-3}$ was modeled as

$$\kappa = aT^{-b} \quad (2)$$

with $a = 272.08$ and $b = 1.1372$ for temperature in K and thermal conductivity in W m$^{-1}$ K$^{-1}$ determined by fitting to existing data.[1] For a 150 nm gap with the maximum emitter temperature of 460 °C and 10 μm, this resulted in 0.76 – 1.13 W or about 19-26% of the total estimated heat transfer to the cell.

**Thermal radiation and photocurrent modeling**

To calculate the spectral and total radiative heat transfer as well as the photocurrent in both the near- and far-field cases, we used a multilayer formalism of fluctuational electrodynamics.[2] This is necessary in the near-field because typical far-field thermal radiation calculations only consider propagating waves that are described by the Planck distribution. In the near-field evanescent waves also contribute significantly to the heat flux via photon tunneling, resulting in much higher heat transfer than predicted by far-field theories.

The fluctuational electrodynamics formalism solves the stochastic Maxwell's equations that incorporate fluctuating thermal currents in a material through the fluctuation-dissipation theorem.[3] In this approach, the radiative heat flux from a source layer $s$ to another layer $l$ in a structure can be written as

$$q_{sl} = \int_0^\infty [\Theta(E, T_s) - \Theta(E, T_l)] \mathcal{T}_{sl}(E) dE \quad (3)$$

where Θ is the mean energy of a Planck oscillator, $\mathcal{T}_{sl}$ is the spectral transmission coefficient from layer $s$ to layer $l$, and $E$ is the photon energy. The transmission coefficient can be expressed as a function of dyadic Green's functions and are determined from a scattering matrix approach.[2] These in turn are dependent on the optical properties and geometry of the layers. The net photocurrent generated in layer $l$ due to light emitted from layer $s$ is similarly obtained by integrating to obtain the photon flux above the bandgap of layer $l$ and multiplying by the electron charge $q$:

$$J_{sl} = q \int_{E_g}^\infty \frac{1}{E} [\Theta(E, T_s) - \Theta(E, T_l)] \mathcal{T}_{sl}(E) dE \quad (4)$$

Here $J_{sl}$ is the current density in layer $l$ due to radiation from layer $s$ and $E_g$ is the bandgap energy of layer $l$.

In this formalism, we model all semiconductor layers as intrinsic GaAs or InAs[4] with added Drude terms similar to the approach taken in prior work[5] for the appropriate n- or p-type carrier concentration[6]. The region above the GaAs emitter is modeled as semi-infinite vacuum, while the emitter and all cell layers are finite with a semi-infinite Au[7] layer under the substrate.

*Additional Supplementary Information*

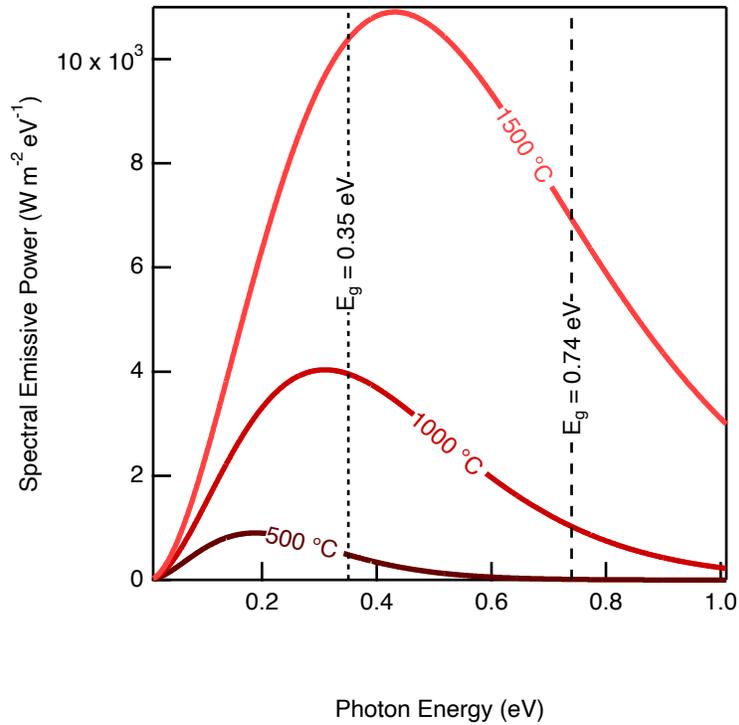

Supplemental Figure 4. Blackbody spectra for temperatures 500 – 1500 °C. The dotted lines indicate the bandgap energy (T= 20 °C) for InAs ($E_g \approx 0.35$ eV) and In$_{.47}$Ga$_{.53}$As ($E_g \approx 0.74$ eV) based cells.

     Figure S4 presents the blackbody spectrum at temperatures between 500 K and 1500 K. The approximate room temperature bandgap energies InAs[8,9] and In$_{.47}$Ga$_{.53}$As[10] are indicated by the dashed and dotted lines respectively. As GaSb[9] and In$_{.47}$Ga$_{.53}$As have nearly identical bandgap energies at room temperature (0.73 eV and 0.74 eV, respectively), only the bandgap energy for In$_{.47}$Ga$_{.53}$As is denoted in Fig. S4. For a blackbody emitter of temperature $T = 500$ °C, the spectral emissive power is more than a order of magnitude higher at the InAs bandgap energy as compared to the In$_{.47}$Ga$_{.53}$As bandgap energy.

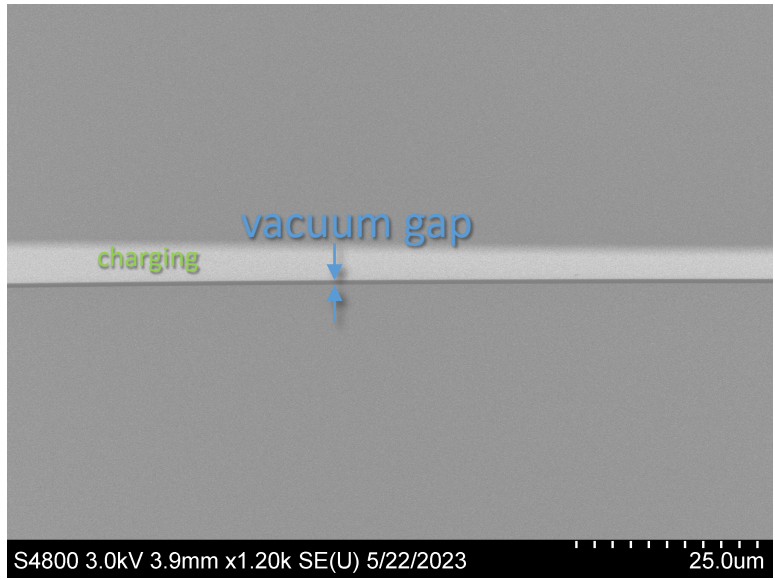

Supplemental Figure 5: Lower magnification scanning electron micrograph of vacuum gap presented in Fig. 2c.

     Figure S5 presents a lower magnification scanning electron micrograph of the vacuum gap presented in Figure 2c. The change in the contrast on the topside is a charging effect near the vacuum gap. The presence of charging effects on a single side of the gap indicates the two sides are not in electrical contact

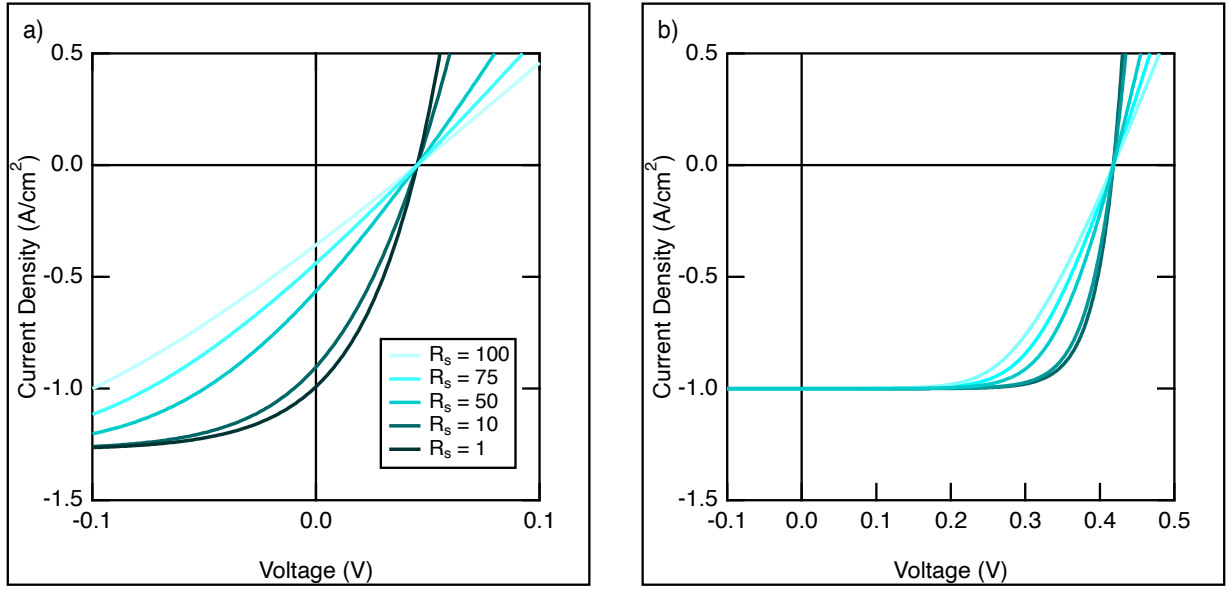

Supplemental Figure 6. Current Density as a function of voltage for a) an InAs based cell and b) an In.53Ga.47As cell with various series resistances (in mΩ•cm²) with nominal cell temperature 335 K and a 1 A/cm² photocurrent.

Figure S6 presents a comparison of the influence of series resistance on an InAs TPV cell (Fig. S6(a)) and In.53Ga.47As TPV cell (Fig. S6(b)). The curves were modeled using the two-diode model presented in equation 5:

$$J = J_{01}\left(\exp\left(\frac{q(V-JR_s)}{k_bT}\right) - 1\right) + J_{02}\left(\exp\left(\frac{q(V-JR_s)}{2k_bT}\right)\right) + \frac{V-JR_s}{R_{sh}} - J_{ph} \quad (5)$$

Where $J$ is the current density in A/cm², $J_{01}$ and $J_{02}$ are the dark current saturation densities in A/cm², $q$ is the elementary charge, $V$ is the voltage, $R_s$ is the series resistance in Ω•cm², $k_b$ is the Boltzmann constant, $T$ is the cell temperature in K, $R_{sh}$ is the shunt resistance in Ω•cm², and $J_{ph}$ is the induced photocurrent in A/cm². The temperature dependence of the two dark current densities are expected to be dominated by the intrinsic carrier concentration[11] (for additional details see ref. 11) leading to the following relationships:

$$J_{01} = C_2 T^3 \exp\left(\frac{-E_g}{k_b T}\right) \quad (6)$$

$$J_{02} = C_2 T^{1.5} \exp\left(\frac{-E_g}{2k_b T}\right) \quad (7)$$

Where $C_1$ and $C_2$ are constants. The temperature dependent bandgap is modeled by the Varshni equation[8]:

$$E_g = E_0 - \frac{\alpha T^2}{(\beta + T)} \quad (8)$$

Where $E_g$ is the temperature dependent bandgap energy in eV, $E_0$ is the bandgap energy at T=0 K in eV, and $\alpha$ and $\beta$ are material specific constants in eV/K and K, respectively. The values used in the modeling are presented in Supplemental Table 1.[8,10]

We set $C_1$ and $C_2$ to be 1 mA/cm². We note that, based on the presence of multiple inflection points in the dark J-V presented in Figure 2(d), our cells may be better modeled by a three-diode model including Auger-Meitner recombination. But these values yield a combined dark current of 14.2 mA/cm² at 283 K and 28.1 mA/cm² at 293 K (Fig. S7). As we measured a dark current of approximately 20 mA at room temperature, we consider these values to be a reasonable estimate. We do not expect Auger-Meitner recombination to be significant in In.53Ga.47As cells and so choose to rely on a two-diode model here for comparison in both cases. We note also that for GaAs cells, Nell reports C = 3.3 mA/cm²•K³ for a single diode model.[12] For consistency, we use these same values for $C_1$ and $C_2$ for the modeled In.53Ga.47As behavior (Fig. S6). Likewise, the temperature is set to 335 K and the photocurrent is set to 1 A/cm² in both cases. In the case of the InAs cell, due to the high dark current, even when the series resistance is very low (1 mΩ•cm²) we observe some deviation between the $J_{sc}$ and $J_{ph}$. As the series resistance increases, this deviation grows extremely quickly. We note also that even under high illumination intensity, in the case of the InAs cells, the high combined dark current at this temperature (modeled at 304.6 mA/cm2) leads to a low open circuit voltage. By comparison, while the In.53Ga.47As cell shows clear evidence of changes to the curve shape as the series resistance increases, $J_{sc} \approx J_{ph}$ in all cases as the combined dark current is four orders of magnitude lower. We note also that, over this full temperature range, In.53Ga.47As is well modeled by a single diode model with an ideality factor of 2 as $J_{0,2\ InGaAs} \gg J_{0,1\ InGaAs}$.

Supplemental Table 1

| Material | $E_0$ (eV) | $\alpha$ (10⁻⁴ eV/K) | $\beta$ (K) | Ref. for all bandgap values | $C_1$ (mA/(cm²•K³)) | $C_2$ (mA/(cm²•K³)) |
|---|---|---|---|---|---|---|
| InAs | 0.426 | 3.158 | 93 | [8] | 1 | 1 |
| In.53Ga.47As | .803 | 4.0 | 226 | [10] | 1 | 1 |

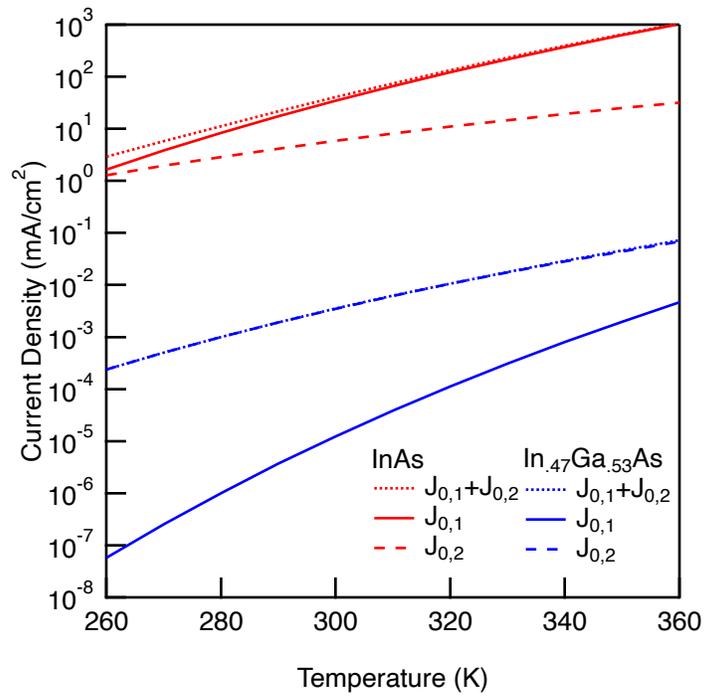

Supplemental Figure 7. Modeled dark current saturation densities for InAs and In$_{.47}$Ga$_{.53}$As as a function of temperature.

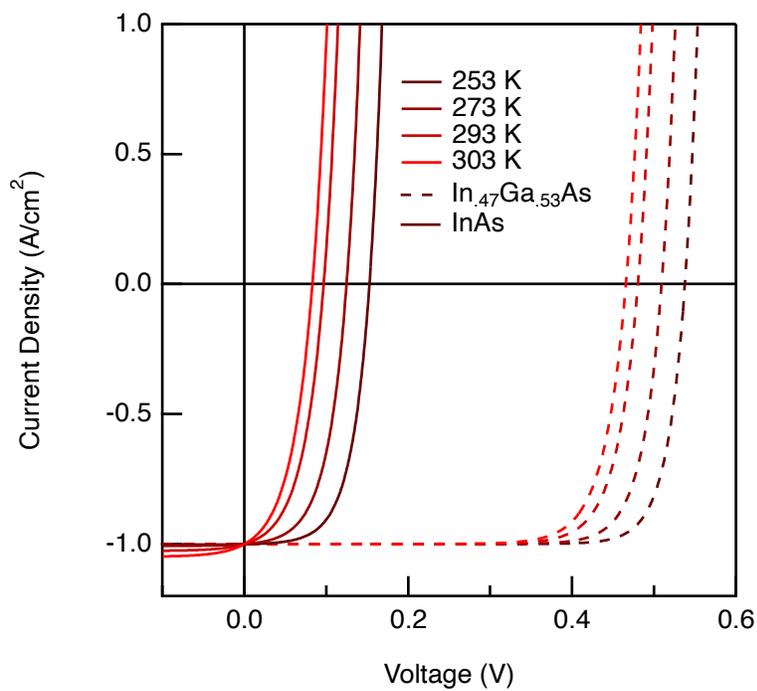

Supplemental Figure 8. Impact of temperature on modeled current density-voltage behavior for $In_{.47}Ga_{.53}As$ and InAs cells, assuming $R_s$ = 0.1 mΩ•cm², $R_{sh}$ = 10000 Ω•cm2, $J_{ph}$ = 1 A/cm² and the values reported in Supplemental Table 1.

     Figure S8 presents the impact of temperature on the diode behavior for $In_{.53}Ga_{.47}As$ and InAs devices with increasing temperature, assuming impacts from series resistance and shunt resistance are negligible. While the decrease in open-circuit voltage is similar in magnitude in both cases, in the case of the InAs cells, this reduction represents a larger proportional change.

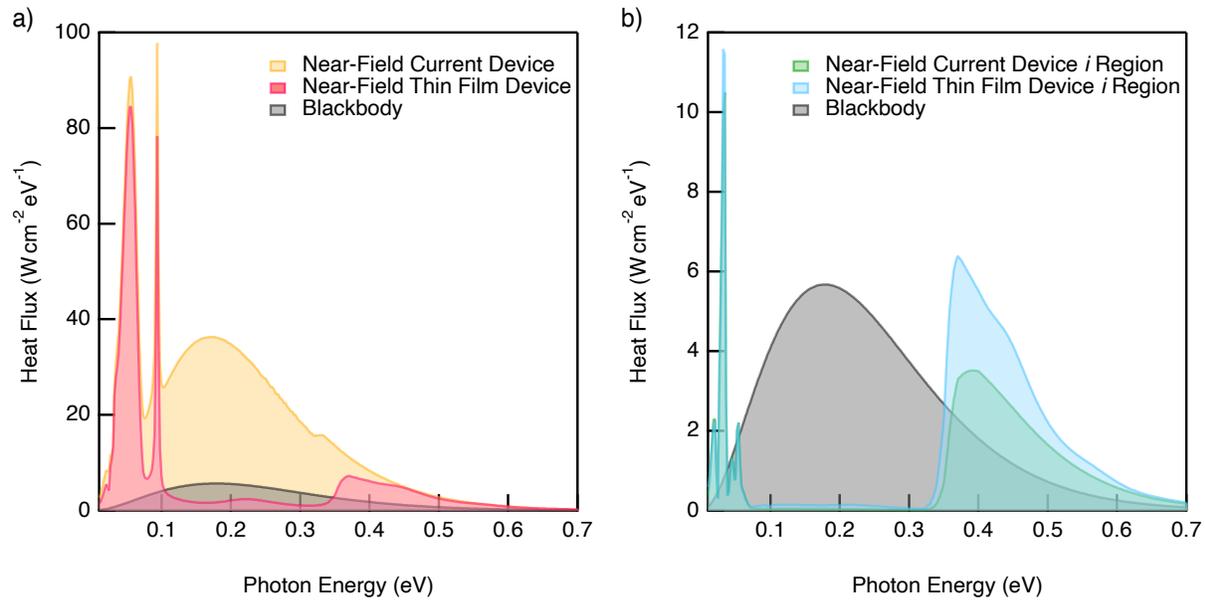

Supplemental Figure 9. Modeled near-field heat flux as compared to the blackbody spectrum at 460 °C for (a) the full cell and (b) the intrinsic layer (used to approximate the minority carrier region). Significant below band gap radiative transfer occurs in the near-field case resulting in cell heating, although the effect is substantially reduced by the thin film geometry. In addition to suppressing some of the sub-bandgap flux, the thin film geometry increases the above bandgap flux as compared to the current device geometry.

Figure S9 presents a comparison of the photon flux assuming a 150 nm gap size, in two near-field cases (the current cell geometry and a thin-film geometry with an Au back reflector) and the far-field cases at 460 °C. The total flux absorbed by the cell in the two near-field cases is presented in Fig S9(a). Sub-bandgap optical flux accounts for the majority of the total energy transfer in all three cases, although the effect is substantially more pronounced in the near-field case with the current cell geometry. As this sub-bandgap light can be absorbed by free-carriers and lead to cell heating, reflecting it should be highly advantageous, as demonstrated by the thin film case with a model Au backside reflector. Nevertheless, the low energy peaks associated with the surface modes remain and finding approaches to reduce these peaks is also extremely important for increasing efficiency and reducing cell cooling requirements. Note also that a rear reflector should also be advantageous in the far-field case at reducing sub-bandgap absorption, however the power generation will remain low.

As only photons absorbed in the minority carrier region generate current, Figure S9(b) presents the optical flux absorbed in the intrinsic region of the device, which should provide a reasonable approximation for the minority carrier region. While both near-field geometries provide an enhancement in above bandgap photon absorption, the effect is substantially larger in the thin film case, as compared to the current geometry, helping to explain the significant power enhancement available in the thin film case.